\documentclass{appolb}
\usepackage{epsfig}
\begin{document}
\title{Collinear improvement of the BFKL kernel in the electroproduction 
of two light vector mesons
\thanks{Presented by F.~Caporale at the ``School on QCD, low-$x$ physics, saturation and diffraction'', Copanello 
(Calabria, Italy), July 1 - 14, 2007.}
}
\author{F.~Caporale, A.~Papa
\address{\it Dipartimento di Fisica, Universit\`a
della Calabria \\
and Istituto Nazionale di Fisica Nucleare, Gruppo collegato di Cosenza \\
I-87036 Arcavacata di Rende, Cosenza, Italy}
\and
A.~Sabio~Vera
\address{\it Physics Department, Theory Division, CERN, CH--1211, Geneva 23, Switzerland}
}

\maketitle
\begin{abstract}
We test the performance of a RG-improved kernel in the determination of the
amplitude of a physical process, the electroproduction of two light vector mesons,
in the BFKL approach at the next-to-leading approximation (NLA). 
We find that a smooth behavior of the amplitude 
with the center-of-mass energy can be achieved, setting the renormalization and 
energy scales appearing in the subleading terms to values much closer to the 
kinematical scales of the process than in approaches based on unimproved kernels.
\end{abstract}
\PACS{12.38.Bx, 13.60.Le, 11.55.Jy}
  
\section{Introduction}
It is well known that the NLA corrections to the BFKL~\cite{BFKL}
Green's function turn out to be very large, this being a signal of the bad behavior 
of the BFKL series. In order to ``cure" the resulting instability, more convergent 
kernels have been introduced, including terms generated by renormalization 
group (RG), or collinear, analysis~\cite{Salam}. They are based on the 
$\omega$-shift method, $\omega$ being the variable Mellin-conjugated 
to the squared center-of-mass energy $s$.
In Ref.~\cite{SabioBess} this original approach has been 
revisited and an approximation to the original $\omega$-shift has been performed,
leading to an explicit expression for the RG-improved NLA kernel.
It would be quite interesting to test the RG-improvement of the kernel 
in the calculation of a full physical amplitude.
A test-field for this comparison can be provided by the physical process 
$\gamma^* \gamma^* \to VV$, where $\gamma^*$ represents a virtual photon and $V$ a 
light neutral vector meson ($\rho^0, \omega, \phi$).
The amplitude of this reaction\footnote{The same process has been analyzed, with different approaches, also in \cite{pire}.} has been calculated in Ref.~\cite{AlexDima1} through 
the convolution of the (unimproved) BFKL Green's function with the $\gamma^* \to 
V$ impact factors, calculated in Ref.~\cite{IKP04}. 
For this amplitude a smooth behavior in $s$ could be achieved by ``optimizing'' the
choice of the energy scale $s_0$ and of the renormalization scale $\mu_R$, which 
appear in the subleading terms. The optimal 
values of the two energy parameters turned out to be quite far from 
the kinematical scales of the reaction, probably because they mimic the unknown next-to-NLA corrections, 
which should be large and of opposite sign respect to the NLA in order to preserve 
the renorm- and energy scale invariance of the exact amplitude.
If this explanation is correct and if the RG-improvement of the kernel catches 
the essential dynamics from subleading orders, then, by the use of an 
RG-improved kernel, one should get more ``natural" values for the optimal choices
of the energy scales and, of course, results consistent with the previous 
determinations.
%In this work we want to address this question, by calculating the NLA amplitude
%of $\gamma^* \gamma^* \to VV$ process in the BFKL approach with the RG-improved 
%kernel of Ref.~\cite{SabioBess}.

\section{The NLA amplitude with the RG-improved Green's function: numerical results}
We consider the production of two light vector mesons ($V=\rho^0, \omega, \phi$) in
the collision of two virtual photons $\gamma^*(Q_1) \: \gamma^*(Q_2)\to V(p_1) \:V(p_2) \;.$
The action of the modified BFKL kernel on his leading eigenfunctions is (the details of all the analytical calculations can be found in Ref.~\cite{Io&Ale}):
\vspace{-0.2cm}
\begin{eqnarray} 
\hat K|\gamma\rangle &=&
\bar \alpha_s(\mu_R) \chi(\gamma)|\gamma\rangle
 +\bar \alpha_s^2(\mu_R)
\left(\chi^{(1)}(\gamma)
+\frac{\beta_0}{4N_c}\chi(\gamma)\ln(\mu^2_R)\right)|\gamma\rangle
\nonumber \\
&+& \bar
\alpha_s^2(\mu_R)\frac{\beta_0}{4N_c}\chi(\gamma)\left(-\frac{\partial}{\partial \gamma}
\right)|\gamma\rangle  + \chi_{RG} (\gamma) |\gamma\rangle \;,
\end{eqnarray} 
where the first term represents the action of LLA kernel, the second
and the third ones stand for the diagonal and the non-diagonal parts of the
NLA BFKL kernel~\cite{AlexDima1} and 
\begin{eqnarray} 
\chi_{RG} (\gamma) &=& 2 \Re e \left\{\sum_{m=0}^{\infty} 
\left[\left(\sum_{n=0}^{\infty}\frac{(-1)^n (2n)!}{2^n n! (n+1)!}
\frac{\left({\bar \alpha}_s+ {\rm a} \,{\bar \alpha}_s^2\right)^{n+1}}
{\left(\gamma + m - {\rm b} \,{\bar \alpha}_s\right)^{2n+1}}\right) 
\right. \right. \\
&-&  \left.\left. \frac{\bar{\alpha}_s}{\gamma + m} - \bar{\alpha}_s^2 
\left(\frac{\rm a}{\gamma +m} + \frac{\rm b}{(\gamma + m)^2}
-\frac{1}{2(\gamma+m)^3}\right)\right]\right\} \nonumber
\end{eqnarray} 
is the solution of the $\omega$-shift equation obtained in \cite{SabioBess}, with 
\begin{eqnarray}
\label{ab}
{\rm a} &=& \frac{5}{12}\frac{\beta_0}{N_c} -\frac{13}{36}\frac{n_f}{N_c^3}
-\frac{55}{36}, \;\;\;\;\;
{\rm b} ~=~ -\frac{1}{8}\frac{\beta_0}{N_c} -\frac{n_f}{6 N_c^3}
-\frac{11}{12}.
\end{eqnarray}
We present our numerical results for the dependence in $s$ of the 
BFKL amplitude calculated for the process under study, using both the 
``exponentiated'' and the ``series'' representations~\cite{AlexDima1}, equivalent within NLA accuracy.
Following Ref.~\cite{AlexDima1}, we will adopt the principle of minimal 
sensitivity (PMS)~\cite{Stevenson} requiring, for each value of $s$, the minimal 
sensitivity of the predictions to the change of both the renormalization and the 
energy scales, $\mu_R$ and $s_0$.

\begin{figure}[tb]
\hspace{-1.3cm}
\begin{minipage}{.48\textwidth} 
\begin{center} {\parbox[t]{4cm}{\epsfysize 4.0cm \epsffile{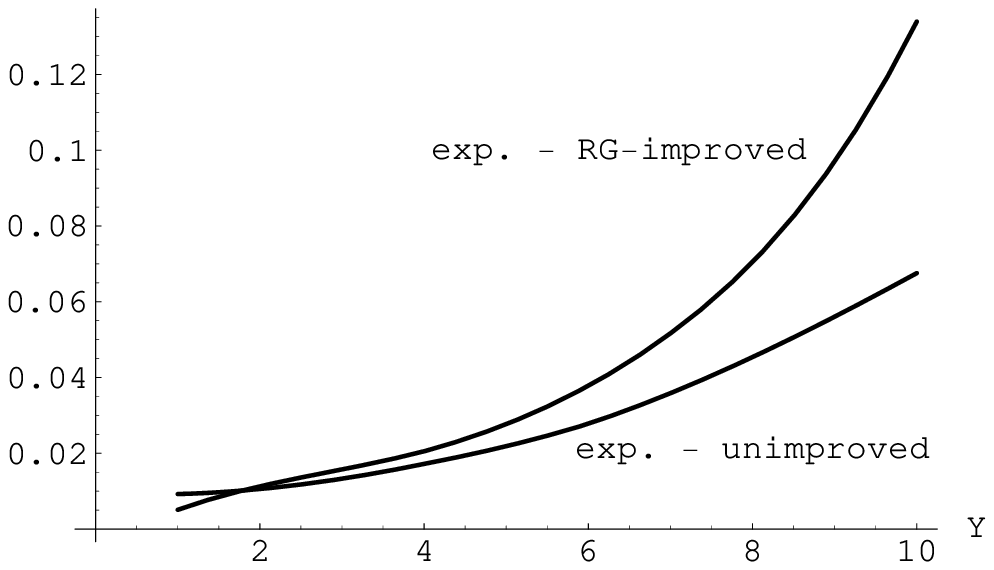}}} 
\end{center} 
\end{minipage}
\hspace{0.45cm}
\begin{minipage}{.48\textwidth} 
\begin{center} {\parbox[t]{4cm}{\epsfysize 4.0cm \epsffile{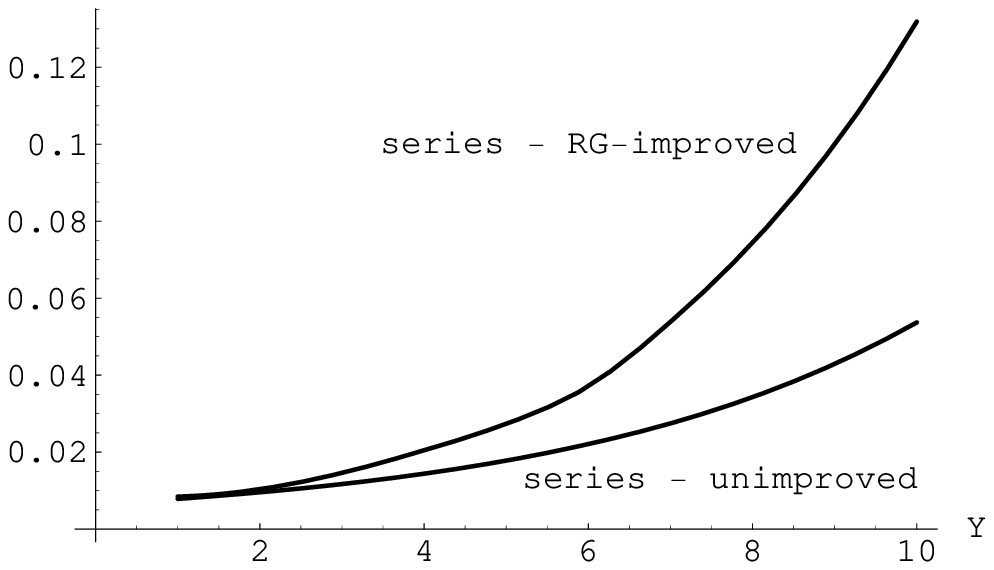}}} 
\end{center} 
\end{minipage}\\
\caption[]{${\cal I}m_s ({\cal A})Q^2/(s \, D_1 D_2)$ as a function of
$Y$ at $Q^2$=24~GeV$^2$ and $n_f=5$ in the ``exponentiated" (left) and ``series" (right) representation 
with and without RG-improvement of the kernel.}
\label{RG}
\vspace{-0.3cm}
\end{figure}
\vspace{-0.2cm}
\subsection{Symmetric kinematics}
We consider here the $Q_1=Q_2\equiv Q$ kinematics, i.e. the ``pure'' BFKL regime, 
with $Q^2$=24~GeV$^2$ and $n_f=5$. We set $\ln(s/s_0)=Y-Y_0$, where $Y=\ln(s/Q^2)$ and $Y_0=\ln(s_0/Q^2)$
and we have looked for the optimal value for the scales $\mu_R$ and $Y_0$. 
We have found that for both representations the amplitude is always quite stable under variation of the 
scales and exhibits generally only one stationary point. We 
choose as optimal values of the parameters those corresponding
to this stationary point.
For the ``exponentiated" representation the optimal values turned out to be typically $\mu_R \simeq 3Q$ and 
$Y_0 \simeq 2$ while for the ``series" representation we have found $\mu_R \simeq 3Q$ and $Y_0 \simeq 3$.
In comparison with Ref.~\cite{AlexDima1}, where the optimal choices 
were typically $Y_0 \simeq 2$ and $\mu_R \simeq 10Q$, we can see that there is a remarkable move
towards ``naturalness''. 
In Fig.~\ref{RG} we show the results for the (imaginary part of the) 
``improved" amplitude in the two representations compared with the result obtained in Ref.~\cite{AlexDima1}. 
Looking at the first plot, the curves are in good agreement at the lower energies, the deviation increasing for 
large values of $Y$. This is consistent with having a larger asymptotic intercept when 
the RG-improvements are taken into account. 
Moreover when the condition $\bar \alpha_s(\mu_R) Y \sim 1$ his satisfied ($Y \sim 6$)
the discrepancy is not so pronounced.
%Moreover the applicability domain of 
%the BFKL approach is determined by the condition $\bar \alpha_s(\mu_R) Y \sim 1$, 
%that for our typical optimal value of $\mu_R$ and for $Q^2$=24~GeV$^2$ means 
%$Y \sim 6$. Around this value the discrepancy is not so pronounced.
In the case of the ``series" representation (Fig.~\ref{RG}, second plot) the situation is similar to the previous one,
but the deviation between the curves appears to be more marked here.
We observe that both the curves for the amplitude with RG-improvement 
fall almost on top of each other.
%while in the 
%determination without the RG-improvement there was a discrepancy, more 
%pronounced at higher energies~\cite{AlexDima1}.
This is a further indication of a better stability, induced by the RG-improvement.
\begin{figure}[tb]
\centering
{\epsfysize 4.4cm \epsffile{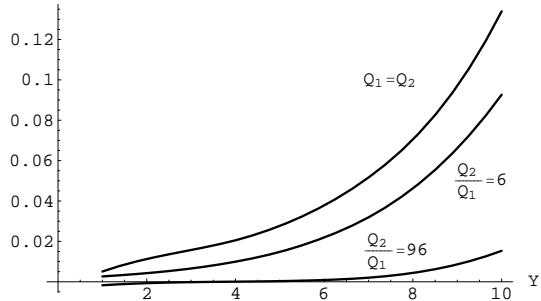}}
\caption[]{${\cal I}m_s ({\cal A})Q_1 Q_2/(s \, D_1 D_2)$ as a function of $Y$ for photons
with strongly ordered virtualities ($Q_2/Q_1=6$ and $Q_2/Q_1=96$, with $Q_1 Q_2$=24~GeV$^2$), 
in comparison with the case of photons with equal virtualities ($Q_1^2=Q_2^2$=24~GeV$^2$).
%All curves have been obtained using the ``exponentiated" representation with the 
%collinearly improved kernel.
}
\label{Asymmtot}
\vspace{-0.3cm}
\end{figure}
\vspace{-0.2cm}
\subsection{Asymmetric kinematics}
When the virtualities of the photons are strongly ordered, we enter 
the ``DGLAP'' regime, where RG-effects should come heavily into the game.
In this regime, previous attempts to numerically determine the amplitude using
unimproved kernels were unsuccessful due to severe instabilities~\cite{AlexDima3}.
We have found here that these instabilities disappear if, instead, the 
RG-improved kernel is used.
In the numerical analysis to follow, we consider two choices for the 
virtualities of the photons, $Q_1$=2~GeV, $Q_2$=12~GeV and $Q_1$=0.5~GeV,
$Q_2$=48~GeV, so that $Q_1 Q_2$ $= Q^2$=24 GeV$^2$ in both cases, and
used the ``exponentiated" representation. We define $Y=\ln(s/Q_1 Q_2)$  
and $Y_0=\ln(s_0/Q_1 Q_2)$. 
For the first choice of virtualities, we find that for each $Y$ value 
the amplitude is still quite stable under variation of the energy parameters and 
the optimal values are $\mu_R \simeq 4\sqrt{Q_1 Q_2}$ and $Y_0 \simeq 2$,
almost independently of $Y$. The same holds for the second choice of virtualities, 
with the only difference that now the optimal values depend strongly on $Y$.
As an example, for $Y=6$, when $\bar \alpha_s(\mu_R) Y \sim 1$, the optimal 
$\mu_R$ is $\simeq 3\sqrt{Q_1 Q_2}$, but $Y_0$=7. This large value for $Y_0$
should not be surprising: if we use $Q_2^2$ as normalization scale in $Y_0$
instead of $Q_1 Q_2$, the optimal value lowers down $\sim $2.5, which looks
more ``natural''.
In Fig.~\ref{Asymmtot} we plot the amplitude for the two choices of 
photons' virtualities we have considered, together with the amplitude for
$Q_1=Q_2=\sqrt{24}$ GeV. The amplitude becomes smaller and smaller when 
$Q_2/Q_1$ increases, as it must be expected due to the presence of the 
factor $\cos(\nu \log(Q_2^2/Q_1^2))$ \cite{Io&Ale} in the integration over $\nu$. 
%We stress
%again that, if the RG-generated terms are removed, it is impossible even to draw
%the curves in Fig.~\ref{Asymmtot} with $Q_2\neq Q_1$.
\vspace{-0.2cm}
\section{Conclusions}
We have applied a RG-improved kernel to determine the amplitude for the forward 
transition from two virtual photons to two light vector mesons in the Regge 
limit of QCD with next-to-leading order accuracy. 
The result obtained is independent on the energy scale $s_0$, and on the
renormalization scale $\mu_R$ within the next-to-leading approximation.
Using two different representations of the amplitude,
we have performed a numerical analysis both in the kinematics of
equal and strongly ordered photons' virtualities.
An optimization procedure, based on the principle of minimal sensitivity,
has led to results stable in the considered energy interval, which allow to 
predict the energy behavior of the forward amplitude. The important finding is that 
the optimal choices of $s_0$ and $\mu_R$ are much closer to the kinematical
scales of the problem than in previous determinations based on unimproved kernels. 
%This effect is very marked for $\mu_R$, as it must be expected, since the extra-terms
%depend on $\mu_R$ and not on $s_0$. 
This leads us to conclude that the extra-terms in the BFKL kernel coming from 
RG-improvement, which are subleading to the NLA, catch an important 
fraction of the dynamics at higher orders.
Moreover, the use of the improved kernel has allowed to obtain the 
energy behavior of the forward amplitude in the case of strongly ordered  
photons' virtualities, which turned out to be unaccessible to previous attempts
using unimproved kernels.


\vspace{-0.2cm}
\begin{thebibliography}{99}
\vspace{-0.2cm}
\bibitem{BFKL}
V.S.~Fadin, E.A.~Kuraev, L.N.~Lipatov, {\it Phys. Lett.} {\bf B60}, 50 (1975);
E.A.~Kuraev, L.N.~Lipatov and V.S.~Fadin, {\it Zh. Eksp. Teor. Fiz.} {\bf 71}, 840 (1976)
[{\it Sov. Phys. JETP} {\bf 44}, 443 (1976)]; {\bf 72}, 377 (1977)  
[{\bf 45}, 199 (1977)];
Ya.Ya.~Balitskii and L.N.~Lipatov, {\it Sov. J. Nucl. Phys.} {\bf 28}, 822 (1978).
\bibitem{Salam}
G.P.~Salam, {\it JHEP} {\bf 9807}, 019 (1998).
\bibitem{SabioBess}
A.~Sabio Vera, {\it Nucl. Phys.} {\bf B722}, 65 (2005).
\bibitem{AlexDima1}
D.Yu.~Ivanov and A.~Papa, {\it Nucl. Phys.} {\bf B732}, 183 (2006); D.Yu. Ivanov and A. Papa, {\it Eur. Phys. J.} {\bf C49}, 947 (2007).
\bibitem{pire} 
B. Pire, L. Szymanowski {\it Eur. Phys. J.} {\bf C44}, 545 (2005); R.~Enberg, B.~Pire and S.~Wallon, { \it  Eur. Phys. J.} {\bf C45}, 759 (2006), {\it Erratum-ibid.} {\bf C51},1015 (2007); B.~Pire, M.~Segond, L.~Szymanowski
{\it Phys. Lett.} {\bf B639}, 642 (2006).
\bibitem{IKP04}  
D.Yu. Ivanov, M.I.~Kotsky and A. Papa, {\it Eur. Phys. J.} {\bf C38}, 195 (2004).
\bibitem{Io&Ale}
F.~Caporale, A.~Papa and A.~Sabio~Vera, {\it Eur. Phys. J.} {\bf C53}, 525 (2008). 
\bibitem{Stevenson}
P.M.~Stevenson, {\it Phys. Lett.} {\bf B100} (1981) 61; {\it Phys. Rev.} {\bf D23}, 2916 (1981).
%\bibitem{AlexDima2}
%D.Yu. Ivanov and A. Papa, {\it Eur. Phys. J.} {\bf C49}, 947 (2007).
\bibitem{AlexDima3}
D.Yu. Ivanov, private communication.
\end{thebibliography}
\end{document}